# Designing a Mobile Game for Home Computer Users to Protect Against "Phishing Attacks"

Nalin Asanka Gamagedara Arachchilage
*School of Information Systems, Computing and Mathematics*
*Brunel University*
*Uxbridge, Middlesex, UK*
Nalin.Asanka@brunel.ac.uk

Melissa Cole
*School of Information Systems, Computing and Mathematics*
*Brunel University*
*Uxbridge, Middlesex, UK*
Melissa.Cole@brunel.ac.uk

### Abstract

*This research aims to design an educational mobile game for home computer users to prevent from phishing attacks. Phishing is an online identity theft which aims to steal sensitive information such as username, password and online banking details from victims. To prevent this, phishing education needs to be considered. Mobile games could facilitate to embed learning in a natural environment. The paper introduces a mobile game design based on a story which is simplifying and exaggerating real life. We use a theoretical model derived from Technology Threat Avoidance Theory (TTAT) to address the game design issues and game design principles were used as a set of guidelines for structuring and presenting information. The overall mobile game design was aimed to enhance avoidance behaviour through motivation of home computer users to protect against phishing threats. The prototype game design is presented on Google App Inventor Emulator. We believe by training home computer users to protect against phishing attacks, would be an aid to enable the cyberspace as a secure environment.*

## 1. Introduction

Home computer users play a significant role in helping to make cyberspace a safer place for everyone due to the internet technology growth. Internet technology is so pervasive today that it provides the backbone for modern living enabling ordinary people to shop, socialize, and be entertained all through their home computers. As people's reliance on the internet grows, so the possibility of hacking and other security breaches increases [10]. Therefore, the message "security begins at home" should be spread to all computer users [3].

Due to the lack of security awareness, professionalism, and training, home computer users create an open back door for hackers using social networking websites [37]. This could be through internet enable services such as Facebook, Twitter, Hi5, Orkut, Skype, and even more professional social networking website like LinkedIn. Therefore, social engineering is still infancy and a constant threat as people give away too much information such as username, password and credit/debit card information through social media.

In addition, as organisations have become increasingly 'virtual' there has been a technological shift from work to the domestic environment [14]. Employees are free to work at home or bring unfinished work home due to the pervasiveness of internet technology. This increases the opportunity for individual users to open themselves to vulnerable IT threats. Unlike employees in the organisations, these home computer users are unlikely to have a sufficient IT infrastructure to protect themselves from malicious IT attacks, or may not have a proper standard or strict IT security policies in place. For example, most home computer users are not IT professionals and lack a high degree of computer literacy to set up a secure home computing system. In addition, home computer users tend to display unsafe computer behaviour which is particularly vulnerable to IT threats. For example, browsing unsafe websites, downloading suspicious software, sharing passwords among family and peers, and using unprotected home wireless networks [10].

Security exploits can include malicious IT threats such as viruses, malicious software (malware), unsolicited e-mail (spam), monitoring software (spyware), the art of human hacking (social engineering) and online identity theft (phishing). One such IT threat that is particularly dangerous to home computer users is phishing. This is a type of semantic attack [2], in which victims get invited by spam emails to visit fraudulent websites. The attacker creates a fraudulent website which has the look-and-feel of the legitimate website. Users are invited by sending emails to access to the fraudulent website and steal their money. It is sometimes much easier and less risky to manipulate human in home setting who have access to sensitive information rather than try and break into systems straight away [37]. For example, spear phishing attack, which is an e-mail spoofing fraud attempt that targets a specific user or an organisation.

Phishing attacks get more sophisticated day by day as and when attackers learn new techniques and change their strategies accordingly [22]. The most popular of which is email [18, 24]. Phishing emails occupy a variety of tactics to trick people to disclose their confidential information such as usernames, passwords, national insurance numbers and credit/debit card numbers. For example, asking people to take part of a survey or urging people to verify their bank account information in which they must provide their bank details to be compensated. The increasing sophistication of these techniques makes individual users difficult to protect from phishing attacks [4]. Automated computer systems can be used to identify some fraudulent emails and websites, but these systems are not totally reliable in detecting phishing attacks [28]. Previous research has revealed that the available anti-phishing tools such as CallingID Toolbar, Cloudmark Anti-Fraud Toolbar, EarthLink Toolbar, Firefox 2, eBay Toolbar, and Netcraft Anti-Phishing Toolbar are insufficient to combat phishing [25, 28]. On one hand, software application designers and developers will continue to improve phishing and spam detection. However, human is the weakest link in information security [6]. On the other hand, human factor risks can mitigate by educating users against phishing threats [25].

Home computer users are susceptible for phishing attacks [31]. This is because the lack of security awareness and sensitive trust decisions that they make during the online activities such as online banking transactions or purchasing items online. To protect individual users against phishing threats, phishing education needs to be considered. Previous studies have reported end-user education as a frequently-recommended approach to countering phishing attacks [32, 33, 34 and 35]. However, this proposed research focus is to enhance user avoidance behaviour through motivation to protect home computer users against phishing attacks.

This research can be accomplished by designing a mobile game as an educational tool to teach home computer users to protect against phishing attacks.

Therefore, it asks the following questions: The first question is how does the systems developer identify which issues the game needs to address? Once the developer has identified the salient issues, they are faced with second question, what principles should guide to structure this information. A theoretical model derived from TTAT uses to address those mobile game design issues and the mobile game design principles were used as guidelines for structuring and presenting information in the game design.

## 2. Theoretical background

"Sometimes a 'friendly' email message tempts recipients to reveal more online than they otherwise would, playing right into the sender's hand [29]". Phishing is one of the rapidly growing online crimes, in which attacker attempt to fraudulently gain sensitive information from a victim by impersonating a trustworthy entity [45]. Victims are lured by an email asking log onto a website that appears legitimate, but it's actually created to steal their usernames and passwords. Online criminals like phishing attacks are easy to do, but hard to stop, because attacks involve social and psychological tools along with technical tools. For example, attacker can send an email insisting victim to logon to his bank website for identity verification. Actually the bank website would be a fake site which has a look and feel as the legitimate bank website. Therefore, attackers may use technical skills to develop a fake website whilst using social engineering techniques for tempting to reveal more online.

Previous research has noticed that technology alone is insufficient to ensure critical IT security issues. So far, there has been little work on end user behaviour of performing security and preventing users from attacks which are imperative to cope up with malicious IT threats such as phishing attacks [3, 5, 9, 11, 12, 16 and 20]. Many discussions have terminated with the conclusion of "if we could only remove the user from the system, we would be able to make it secure" [27]. Where it is not possible to completely eliminate the user, for example in home use, the best possible approach for computer security is to educate the user in security prevention [17, 2]. However, most home computer users are lack of security awareness due to the deficiency of education, training and professionalism [36]. Previous research has revealed well designed user security education can be effective [28, 22]. This could be web-based training materials, contextual training, and embedded training to improve users' ability to avoid phishing attacks. One objective of our research is to find effective ways to educate people to identify and prevent phishing websites.

There has been previous research on phishing falls into three categories: research related to understand why people fall for phishing attacks, what tools to protect people against phishing attacks, and methods for educating people not to fall for phishing attacks [28, 31, and 35].

Why People fall for phishing attacks? Dhamija et al conducted a laboratory based experiment showing twenty two participants to twenty websites and asked them to determine which were fraudulent [39]. They found out that participants made mistakes on the test set 40% of the time. Furthermore, they noted that 23% of their participants ignored all cues in the web browser address bar, and status bar as well as all security indicators.

Downs et al have studied role playing study aimed at understanding why people fall for phishing emails and what cues they look for to avoid such attacks [38]. Their finding highlighted two key things in their work. First, even people are aware of phishing attacks, they do not think that their own vulnerability or strategies for tracing phishing attacks. Second, while people can secure themselves from known risks, they tempt to have difficulties generalising their known to unfamiliar risks.

What tools to protect people against phishing attacks? Anti-phishing tools are now provided by internet service providers [28]. For example, internet service providers built into mail servers and clients, and available as web browser toolbars. There are currently available anti-phishing tools which are free to download [40]. CallingID Toolbar, Cloudmark Anti-Fraud Toolbar, EarthLink Toolbar, Firefox 2, eBay Toolbar, and Netcraft Anti-Phishing are few examples. These tools do not effectively protect against all phishing attacks [25, 28].

Zhang et al have developed an automated test bed for testing available anti-phishing tools [40]. They used 200 verified phishing URLs and 516 legitimate URLs to test the effectiveness of 10 popular anti-phishing tools. Their findings said, only one tool was able to consistently identify more than 90% of phishing URLs correctly. It also incorrectly identified 42% of legitimate URLs as phish. Other tools varied considerably depending on the source of phishing URLs. Of these remaining tools, only one correctly identified over 60% of phishing URLs. However, it is interesting to know that attackers and tool developers are continuously on an arms race.

Is anti-phishing training or education effective? Even though there are usability and security experts claim that user education and training does not work [28, 41], previous research revealed that well designed end-user education could be a recommended approach to combating phishing attacks [28, 32, 33, 34 and 35].

Kumaraguru et al have designed and evaluated an embedded training email system that teaches protecting people from phishing attacks during their normal use of email [22]. The authors conducted lab experiments contrasting the effectiveness of standard security notices about phishing with two embedded training designs they developed. Finally, they found that embedded training works better than the current practice of sending security notices.

Sheng et al have designed Anti-Phishing Phil: online game that teaches users to avoid phishing threats [28]. They evaluated the game through a user study. Participants were tested on their ability to identify fake websites before and after engaging 15minutes one of three anti-phishing training activities such as playing the game, reading an anti-phishing tutorial or reading existing online training materials. Finally, they found that the participant who played the game were better able to identify fake websites compared to the others. They also confirmed that games can be effective way of educating people avoid phishing attacks.

So how to educate the home user in order to protect from phishing attacks? To address the problem, this research focuses on the design of a mobile game that teaches home computer users to prevent from phishing attacks. This concept is based on the notion that not only can a computer game provide an education [8], but also games potentially provide a better learning environment, because they motivate the user and keep attention by providing immediate feedback [1, 19]. Still, the mobile game indicates its potential to improve learning. The game at least moves users into a different state where they are mentally ready for learning by creating right environment, where they also already experience some social oriented learning [7]. Furthermore, designing a game for mobile devices such as mobile telephones, pagers and personal digital assistants enables individual users to download, install and use conveniently. The most significant feature of mobile environment is mobility itself such as mobility of the user, mobility of the device, and mobility of the service [7]. It enables users to be in contact while they are outside the reach of traditional communicational spaces. For example, a person can play a game on his mobile device while travelling on the bus or train.

## 3. Game design issues

The aim of the proposed game design is to educate home computer users to thwart phishing attacks. To answer this question issues drawn from the threat of threat avoidance will be used to explore the principles needed for structuring the design of the game in the context of personal computer use. The TTAT describes individual IT users' behaviour of avoiding the threat of malicious information technologies such as phishing attacks (Fig. 1) [9]. The model examines how individuals avoid IT threats by using a given safeguarding measure. The safeguarding measure does not necessarily have to be an IT source such as anti-phishing tools; rather it could be behaviour [10].

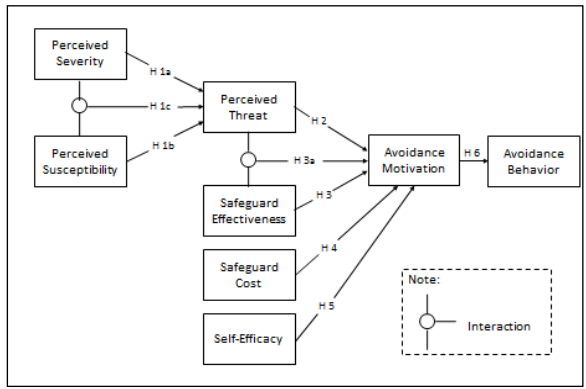

Figure 1. Research model derived from TTAT [10]

Consistent with TTAT [9], user's IT threat avoidance behaviour is determined by avoidance motivation, which, in turn, is effected by perceived threat. Perceived threat is influenced by perceived severity and susceptibility as well as their interaction. User's avoidance motivation is also determined by the three constructs such as safeguard effectiveness, safeguard cost, and self-efficacy. In addition, the research model posits that avoidance motivation is influenced by an interaction between perceived threat and safeguard effectiveness.

Whilst the TTAT forms the issues that the game design needs to address, it should also indicate how to structure this information and present in a mobile device. The proposed mobile game design attempts to develop threat perceptions which the individuals will be more motivated to avoid phishing attacks and use safeguarding measures. They realise the effectiveness of safeguarding measures, lower safeguard costs, and increase self-efficacy.

This game design emphasises both the likelihood of phishing attacks and the severity of losses caused by the attacks. Furthermore, this game design concerned to possible emotion-focus coping and help people understand how they unintentionally engage in emotion-focused coping and how to stop which help people focus on problem-focused coping rather than reducing the negative effects of high threats perceptions [10].

## 3.1 What to teach?

The possible phishing attacks can be identified in several ways, such as carefully looking at the website address, so called Universal Resource Locator (URL), signs and the content of web page, the lock icons and jargons of the webpage, the context of the email message, and the general warning messages displayed in the website [15, 21]. Previous research has identified that existing anti-phishing techniques based on URLs and email messages are not robust enough for phishing detection [15, 26].

Our objective of the anti-phishing mobile game design is to teach user two things. First, how to identify phishing website addresses (URLs). Second, how to identify phishing emails by analysing the content of email message. Therefore, the game design should develop the awareness of identifying the features of website addresses and email messages. For example, legitimate websites usually do not have numbers at the beginning of their URLs such as http://81.153.192.106/.www.hsbc.co.uk or phishing emails often contain generic salutation such as "Dear customer of The Royal Bank of Scotland".

## 3.2 Story

The mobile game design prototype contains two sections: how to identify phishing website addresses (URLs) and phishing emails. The game is based on a scenario of a character of a small fish and 'his' teacher who live in a big pond.

### 3.2.1 How to identify phishing website addresses (URLs)? (Fig. 2)

The main character of the game is the small fish, who wants to eat worms to become a big fish. However he should be careful of phishers those who try to trick him with fake worms. This represents phishing attacks by developing threat perception. Each worm is associated with a website address, so called Unified Resource Locator (URL) which appears as a dialog box. The small fish's job is to eat all the real worms which associate legitimate website addresses and reject fake worms which associate with fake website addresses before the time is up. This attempts to develop the severity and susceptibility of the phishing threat in the game design.

The other character is the small fish's teacher, who is a matured and experienced fish in the pond. If the worm associated with the URL is suspicious and if it is difficult to identify, the small fish can go to 'his' teacher and request a help. The teacher could help him by giving some tips on how to identify bad worms. For example, "website addresses associate with numbers in the front are generally scams," or "a company name followed by a hyphen in a URL is generally a scam". Whenever the small fish requests a help from the teacher, the score will be reduced by certain amount as a payback for safeguard measure. This attempts to address the safeguard effectiveness and the cost needs to pay for the safeguard in the game design.

### 3.2.1 How to identify phishing emails? (Fig. 3)

The main character of the game is the small fish, who wants to eat worms to become a big coloured fish. However he should be careful of phishers those who try to trick him with fake worms. This represents phishing attacks by developing threat perception. Each worm is associated with an email which could be legitimate or fake. The small fish's job is to eat all the real worms which associate legitimate emails while rejecting fake worms which associate fake emails before the time is up. This attempts to develop the severity and susceptibility of the phishing attack in the game design. The content of an email helps user to identify a phishing attack. For example, phishing emails often contain generic salutation or use of a trusted company logo. It could also be a statement using immediate action. The user is presented a worm which associate with an email that may contain phishing email traps as well as legitimate emails. The phishing email traps covered in the game design including fake links, malicious attachments, and much more.

The other character is the small fish's teacher, who is a matured and experienced fish in the pond. If the worm associated with the email is suspicious and if it is difficult to identify, the small fish can go to 'his' teacher and request a help. The teacher could help him by giving some tips on how to identify phishing emails. For example, "phishing emails often contain generic salutation" or "emails associated with urgent requests are generally phishing emails". Whenever the small fish requests a help from the teacher, the score will be reduced by certain amount as a payback for safeguard measure. This attempts to address the safeguard effectiveness and the cost needs to pay for the safeguard in the game design.

The proposed overall game design is presented in different levels such as beginner, intermediate and advance. When the user come across from the beginner to advance level, complexity of the combination of URLs and emails is dramatically increased while considerably decreasing the time period to complete the game. Therefore self-efficacy of preventing from phishing attacks will be addressed in the game design as the user come across from the beginner to advance level. The overall game design is used to increase users' avoidance behaviour through motivation to protect against phishing attacks.

## 4. Game design principles

The above mentioned scenario should combine with a set of guidelines that focused on designing of the educational game [1, 19]. This set of guidelines which is known as game design principles, describes how the user interacts with mobile game. Prensky [19] has proposed that the mobile game can be described in terms of six structural elements. Those elements were used in our game design as guidelines for structuring and presenting information.

1) Rules: which organize the game. The story we developed based on the theoretical model derived from TTAT in our game design describes the rules.

2) Goals and objective: which the players struggle to achieve. The user has the goal to solve the task. This is designed in our game to accomplish all the levels such as beginner, intermediate, and advance by eating real worms (associated with legitimate URLs or emails) while avoiding fake worms (associated with fake URLs or emails).

3) Outcome and feedback: which measure the progress against the goals. The user gets the real time feedback on the current status in the game. For instance, when the small fish eats a real worm which is associated with legitimate URL or email, he might get the real time feedback saying "WOW well done!".

4) Conflict, competition, challenge, and opposition leading to players' excitement: which are realised through the opportunity to gain points against given lives of the small fish in our mobile game design.

5) Interaction: This is known the social aspect in the game. This is accomplished by providing real time feedback, fantasy and rewords or gaining points in our game design. By creating attractive digital objects such as small fish, worms, pond in our game design, not only immerse in an augmented physical environment, but also immerse into an augmented social environment.

6) Representation or story: This is exaggerating interesting aspects of reality. The representation is realized through the scenario or the story we developed using digital objects such as small fish, 'his' teacher, and worms in our game design.

Furthermore, the work not only focuses on design of a mobile game [30], but also it should focus on the usability of the game application [13]. Usability of the game design often attempts to recreate a typical playing environment, to emulate how a player would typically play the game in the real world while allowing the game players to achieve their goals. For example, they should notice when something in the game changes such as decreasing the time period while intricate the combination of URLs or emails when they move to next level or move the digital objects in the game using keys in the keypad, touch stick or finger without any difficulty.

## 5. Game prototype design

To explore the viability of using a game to prevent from phishing attacks, a prototype model was developed for a mobile telephone using Google App Inventor Emulator (Fig. 2 and Fig. 3).

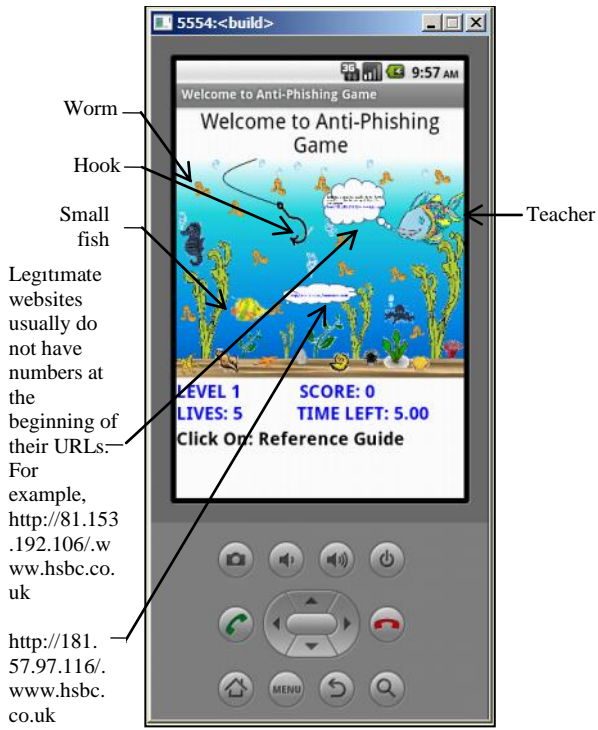

Figure 2. Phishing URLs menu of the game prototype design displayed on Google App Inventor Emulator.

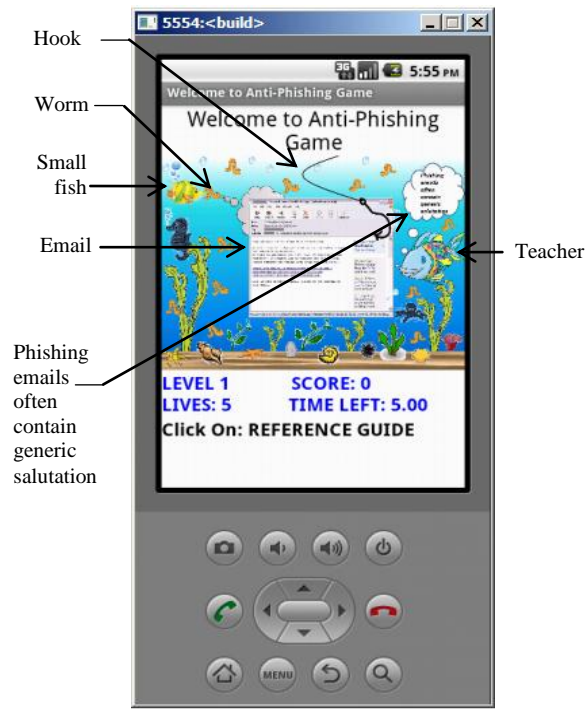

Figure 3. Phishing emails menu of the game prototype design displayed on Google App Inventor Emulator.

The game designed was focused to provide home computer user with encapsulated resources that teaches how to prevent from phishing attacks. This is categorised into two parts such as the game and reference guide. The game design addresses the issues derived from the TTAT and the reference guide provides useful information which the user can learn more about phishing attacks.

The reference guide is linked to the phishing education section of the Anti-Phishing Work Group website (APWG - http://education.apwg.org/) for learn more about phishing education. For example, what is phishing threat, how it could be severe, what is the usefulness of having a safeguarding measure, where to report a suspected phishing email or website, and phishing education to thwart phishing attacks.

## 6. Conclusion

This research focuses on design a mobile game as a tool to educate home computer users by creating awareness of phishing attacks. It addressed two questions: The first question is how does the systems developer identify which issues the game need to address? Then they are faced with second question, what principal should guide them to structure this information? We used a theoretical model derived from TTAT in order to address those issues and mobile game design principles were used as a set of guidelines for structuring and presenting information in the game design. The objective of our anti-phishing mobile game design was to teach user two things: which are how to identify phishing website addresses (URLs) and phishing emails. Those are couple of many ways to identify a phishing attack. Furthermore, as future research we attempt to design a mobile game to teach the other areas such as signs and content of the web page, the lock icons and jargons of the webpage, and the general warning messages displayed in the website. The overall game design was targeted to enhance avoidance behaviour through motivation to protect home computer users from phishing attacks. We believe by educating home computer users against malicious IT threats, would make a considerable contribution to enable the cyberspace as a secure environment.

## 7. References

[1] A. Amory and R. Seagram, "Educational Game Models: Conceptualization and Evaluation", South African Journal of Higher Education, vol. 17 (2), pp. 206-217, 2003.

[2] B. Schneier, "Semantic Attacks, The Third Wave of Network Attacks", Crypto-Gram Newsletter, October 2000, Retrieved from http://www.schneier.com/crypto-gram-0010.html. (Accessed date: 02 April 2011.)

[3] B. Y. Ng and M. A. Rahim, "A Socio-Behavioral Study of Home Computer Users' Intention to Practice Security",

The Ninth Pacific Asia Conference on Information Systems, Bangkok, Thailand, 2005.

[4] C. E. Drake, J. J. Oliver and E. J. Koontz, "Mail Frontier Anatomy of a Phishing Email", February 2006, Retrieved from http://www.mailfrontier.com/docs/MF_Phish_Anatomy.pdf. (Accessed date: 03 April 2011.)

[5] C. L. Anderson and R. Agarwal, "Practicing Safe Computing: Message Framing, Self-View, and Home Computer User Security Behaviour Intentions", International Conference on Information Systems, Milwaukee, WI, pp. 1543-1561, 2006.

[6] CNN. com, "A convicted hacker debunks some myths", 2005, http://www.cnn.com/2005/TECH/internet/10/07/kevin.mitnick.cnna/index.html. (Accessed date: 04 April 2011.)

[7] D. Parsons, H. Ryu and M. Cranshaw, "A Study of Design Requirements for Mobile Learning Environments", Proceedings of the Sixth IEEE International Conference on Advanced Learning Technologies, pp. 96-100, 2006.

[8] E. M. Raybourn and A. Waern, "Social Learning Through Gaming", Proceedings of CHI 2004, Vienna, Austria, pp. 1733-1734, 2004.

[9] H. Liang and Y. Xue, "Avoidance of Information Technology Threats: A Theoretical Perspective", MIS Quarterly, vol. 33 (1), pp. 71-90, 2009.

[10] H. Liang and Y. Xue, "Understanding Security Behaviours in Personal Computer Usage: A Threat Avoidance Perspective", Journal of the Association for Information Systems, vol. 11 (7), pp. 394-413, July 2010.

[11] H. Susan, A. Catherine and A. Ritu, "Practicing Safe Computing: Message Freaming, Self-View and Home Computer User Security Behaviour Intentions", ICIS 2006 Proceedings, pp. 93, 2006, Retrieved from http://aisel.aisnet.org/icis2006/93. (Accessed date: 15 March 2011.)

[12] I. Woon, G. W. Tan and R. Low, "A Protection Motivation Theory Approach to Home Wireless Security", International Conference on Information Systems, Las Vegas, NV., pp. 367-380, 2005.

[13] J. Gong and P. Tarasewich, "Guidelines for Handheld Mobile Device Interface Design", Proceedings of Decision Sciences Institute, Annual Meeting, Boston, Massachusetts, pp. 3751-3756, 2004.

[14] J. O'Brien, T. Rodden, M. Rouncefield and J. Hughes, "At Home with the Technology: An Ethnographic Study of a Set-Top-Box Trial", ACM Transactions on Computer-Human Interaction, vol. 6 (3), pp.282-308, 1999.

[15] J. S. Downs, M. Holbrook and L. F. Cranor, "Behavioural response to phishing risk", Proceedings of the anti-phishing working groups - 2nd annual eCrime researchers summit, pp.37-44, October 2007, Pittsburgh, Pennsylvania, Retrieved from doi>10.1145/1299015.1299019. (Accessed date: 25 March 2011)

[16] K. Aytes and C. Terry, "Computer Security and Risky Computing Practices: A Rational Choice Perspective", Journal of Organizational and End User Computing, vol. 16 (2), pp. 22-40, 2004.

[17] K. Mitnick and W. L. Simon, "The art of Deception – Controlling the Human Elements of Security", 2002.

[18] L. James, "Phishing Exposed", Syngress, Canada, 2005.

[19] M. Prensky, "Digital Game-Based Learning Revolution", Digital Game-Based Learning, New York, 2001.

[20] M. Workman, W. H. Bommer and D. Straub, "Security Lapses and the Omission of Information Security Measures: A Threat Control Model and Empirical Test", Computers in Human Behaviour, vol. 24 (6), pp. 2799-2816, 2008.

[21] M. Wu, R. Miller, and S. Garfinkel, "Do Security Toolbars Actually Prevent Phishing Attacks?", Posters SOUPS, 2005.

[22] P. Kumaraguru, Y. Rhee, A. Acquisti, L. F. Cranor, J. Hong and E. Nunge, "Protecting people from phishing: the design and evaluation of an embedded training email system", Proceedings of the SIGCHI conference on Human Factors in Computing Systems, San Jose, California, USA, April - May 2007.

[23] R. G. Brody, E. Mulig and V. Kimball, "Phishing, pharming and identity theft", Journal of Academy of Accounting and Financial Studies, vol. 11, pp. 43-56, 2007.

[24] R. Richmond, "Hackers set up attacks on home PCs, Financial Firms: study", September 2006, Retrieved from http://www.marketwatch.com/News/Story/Story.aspx?dist=new sfinder&siteid=google&guid=%7B92615073-95B6-452EA3B9 569BEACF91E8%7D&keyword=. (Accessed date: 27 March 2011.)

25] S. A. Robila, J. W. Ragucci, "Don't be a phish: steps in user education", Proceedings of the 11th annual SIGCSE conference on Innovation and technology in computer science education, 26 – 28 June 2006, Bologna, Italy, Retrieved from doi>10.1145/1140124.1140187. (Accessed date: 29 March 2011.)

[26] S. Garera, N. Provos, M. Chew, A. D. Rubin, "A framework for detection and measurement of phishing attacks", Proceedings of the 2007 ACM workshop on Recurring malcode, Alexandria, Virginia, USA, November 2007.

[27] S. Gorling, "The Myth of User Education", Proceedings of the 16th Virus Bulletin International Conference, Royal Institute of Technology, Department of Industrial Economics and Management (INDEK), 2006.

[28] S. Sheng, B. Magnien, P. Kumaraguru, A. Acquisti, L. F. Cranor, J. Hong and E. Nunge, "Anti-Phishing Phil: the design and evaluation of a game that teaches people not to fall for phish", Proceedings of the 3rd symposium on Usable privacy and security, Pittsburgh, Pennsylvania, July 2007.

[29] T. Jagatic, N. Johnson, M. Jakobsson and F. Menczer, "Social Phishing", Communications of the ACM, vol. 50 (10), pp. 94-100, 2007.

[30] X. Korhonen and X. Koivisto, "Mobile Entertainment: Playability Heuristics for Mobile Games", Proceedings of Mobile HCI, 2006.

[31] K. Ponnurangam, Y. Rhee, S. Sheng, S. Hasan, A. Acquisti, L. F. Cranor and J. Hong, "Getting Users to Pay Attention to Anti-Phishing Education: Evaluation of Retention and Transfer", APWG eCrime Researchers Summit, October,4-5, Pittsburgh, PA, USA, 2007.

[32] M. Allen, Social Engineering: A means to violate a computer system. Tech. rep., SANS Institute, 2006.

[33] J. Hiner, Change your company's culture to combat social engineering attacks, November 2002, Retrieved from http://articles.techrepublic.com.com/5100-1035_11-1047991.html. (Accessed date: 15 July 2011.)

[34] D. Timko, The social engineering threat, Information Systems Security Association Journal, 2008.

[35] P. Kumaraguru, S. Sheng, A. Acquisti, L. F. Cranor and J. Hong, "Lessons from a real world evaluation of anti-phishing training", eCrime Researchers Summit, pp. 1–12, 2008. [38] C. P. Hui, How to Study Home Users, TKK T-110.5190, Seminar on Internetworking, 2007.

[36] C. P. Hui, How to Study Home Users, TKK T-110.5190, Seminar on Internetworking, 2007.

[37] Michael Pike, The Magazine for the IT Professional, British Computer Society, The Charted Institute for IT, March 2011.

[38] Downs, J., M. Holbrook and L. Cranor. 2006. Decision strategies and susceptibility to phishing. In Proceedings of the Second Symposium on Usable Privacy and Security (Pittsburgh, Pennsylvania, July 12 - 14, 2006). SOUPS '06, vol. 149. ACM Press, New York, NY, 79-90.DOI= http://doi.acm.org/10.1145/1143120.1143131. (Accessed date : 21 July 2011.)

[39] Dhamija, R., J. D. Tygar. and M. Hearst. 2006. Why phishing works. In Proceedings of the SIGCHI Conference on Human Factors in Computing Systems (Montréal, Québec, Canada, April 22 - 27, 2006). R. Grinter, T. Rodden, P. Aoki, E. Cutrell, R. Jeffries, and G. Olson, Eds. CHI '06. ACM Press, New York, NY, 581-590. DOI= http://doi.acm.org/10.1145/1124772.1124861. (Accessed date: 02 August 2011.)

[40] Zhang, Y., Egelman, S., Cranor, L. F., and Hong, J. Phinding phish - Evaluating anti-phishing tools. In Proceedings of the 14th Annual Network & Distributed System Security Symposium (NDSS 2007) (28th February - 2nd March, 2007). http://lorrie.cranor.org/pubs/toolbars.html. (Accessed date: 01 August 2011.)

[41] Evers, J. Security Expert: User education is pointless. Retrieved, Jan 13, 2007, http://news.com.com/2100-7350_3-6125213.html. (Accessed date: 01 August 2011.)